\documentclass[showpacs,showkeys,12pt,preprint,preprintnumbers,nofootinbib,
groupedaddress,superscriptaddress,amsmath,amssymb]{revtex4}
\usepackage[dvips]{color}
\usepackage{graphicx}

\begin{document}
\title{Effects of the anomalous Higgs couplings
on the Higgs boson production at the Large Hadron Collider}
\author{Shinya Kanemura}
\email{kanemu@sci.u-toyama.ac.jp} \affiliation{Department of
Physics, University of Toyama, 3190 Gofuku, Toyama 930-8555,
Japan}
\author{Koji Tsumura}
\email{ktsumura@ictp.it} \affiliation{The Abdus Salam ICTP of
UNESCO and IAEA, Strada Costiera 11, 34151 Trieste, Italy}
\preprint{UT-HET 015, IC/2008/078}
\pacs{
14.65.Ha, 	
12.60.Fr 	
}
\keywords{Higgs boson, Higher dimensional operator}
\begin{abstract}
We study the impact of dimension-six operators on single- and
double-Higgs production rates via gluon fusion at the Large Hadron
Collider (LHC).
If the top-Yukawa coupling is modified by some new physics whose
scale is of the TeV scale, its
effect changes the cross sections of single-Higgs production
$gg\to H$ and double-Higgs production $gg\to HH$ through
the top-loop diagram. In particular, double-Higgs production
can receive significant enhancement
from the effective top-Yukawa coupling and the new dimension-five
coupling $t{\bar t}HH$ which are induced by the dimension-six
operator. Comparing these results to the forthcoming data at the
LHC, one can extract information of the dimension-six operators
relevant to the top quark and the Higgs boson.
\end{abstract}
\maketitle
\section{Introduction}
\label{sec:introduction}
Gauge symmetries of the standard model (SM) have been well
confirmed at the CERN Large Electron Positron collider
(LEP)~\cite{:2004qh} and the Fermilab TEVATRON.
However the mechanism of spontaneous symmetry breaking has not been tested
yet~\cite{Amsler:2008zz,Barate:2003sz}.
The vacuum expectation value of the Higgs boson triggers
electroweak symmetry breaking and generates masses of weak gauge
bosons, quarks and charged leptons. Search for the Higgs boson is
the main purpose of the measurement at the CERN Large Hadron Collider (LHC).

In the SM, coupling constants of the Higgs boson with the weak
gauge bosons and the matter fields directly relate to their
masses. In order to clarify the mass generation mechanism,
an independent determination of the particle masses and their couplings
to the Higgs boson is important, which will be the subsequent task at the LHC. It
will give not only a confirmation of the SM but also an indication
of new physics beyond the SM.

At the LHC the dominant production mechanism of the Higgs bosons
is gluon fusion $gg\to H$.
The leading contribution to this process comes from top-quark
loop diagrams. Information of the top-Yukawa coupling can be
extracted through this process as a combination with the Higgs decay
branching ratios.
The gauge interaction of the Higgs boson would be tested through
processes of
vector boson fusion $VV^*\to H(V=W^-,Z)$~\cite{Cahn:1983ip,Hankele:2006ma}
and Higgs-strahlung $q{\bar q'}\to VH$~\cite{Glashow:1978ab}.
These processes are promising channels for Higgs searches
too because of the kinematic advantage in the reconstruction of signals.
Measurement of the triple-Higgs boson coupling has been discussed
in the double-Higgs production mechanism from gluon fusion, $gg\to
HH$, at the LHC~\cite{Glover:1987nx}.
In Ref.~\cite{Baur:2003gp} the sensitivity to the Higgs boson
self-coupling is studied. The authors of this reference 
conclude that its experimental
accuracy could reach $20$--$30$\% at the SLHC with an integrated
luminosity of $L=3000$ fb$^{-1}$ for $m_H^{}=150$--$200$ GeV.

Measuring the top-Yukawa coupling accurately
is important because the magnitude of the coupling constant
$(y_t^\text{SM} \sim 1)$ indicates that the physics of top quarks
closely would relate to that of electroweak symmetry breaking.
Lots of models are proposed in this direction~\cite{topmode}.
Measurements of the top-Yukawa coupling would be a key to
uncover such possibilities. In addition,
the Higgs boson self-coupling is of great interest by itself
to understand the nature of spontaneous symmetry breaking.
Its measurement can also be a probe of the new physics beyond the SM.
The coupling strength is also important being deeply related to
the condition of successful electroweak baryogenesis\cite{baryon}.

New physics beyond the SM will be recognized by the discovery of
non-SM particles or by detecting the deviation from the SM
relations between masses and coupling constants. For the latter case,
the low energy effective theory at the electroweak scale can be described by
the SM Lagrangian with additional higher dimensional operators.
This approach has been investigated to analyze non-standard interactions
in a model independent way. Leading order contributions of such non-standard
interactions would be described by the dimension-six operators~\cite{buchmuller}.
Constraints on these operators and their phenomenology have
been discussed in the literature~\cite{Hagiwara-Hatsukano-Ishihara-Szalapski,
Kanemura:2006bh,Barger:2003rs}.

In this paper, we study new physics effects from dimension-six
operators on single- and double-Higgs production processes, $gg\to
H$ and $gg\to HH$. The dimension-six operators correct to the
top-Yukawa coupling and the triple-Higgs boson coupling,
also induce the tree level $ggH$ and $ggHH$ vertices.
Effects due to the modified top-Yukawa coupling and the tree level
coupling on the effective $ggH$ vertex are investigated.
The former comes from color blind new dynamics while the latter
can come from some color dependent effects.
The experimental limits from the LEP precision data and the theoretical
bounds such as the unitarity bounds are taken into
account. We find that the effects on these processes due to the
dimension-six operators can be significant even under these
constraints.
In particular the double-Higgs production cross section is
sensitive to these dimension-six operators. These contributions
from the dimension-six operators can be distinguished by comparing
the data for these Higgs boson production channels at the LHC experiments.

This paper is organized as follows. In Sec. II, we introduce the
dimension-six operators as a new physics effect. Its experimental
and theoretical bounds are discussed. In Sec. III, numerical
evaluations of the effects of dimension-six operators on the Higgs
production processes at the LHC are shown. Conclusions and
discussions are given in Sec. IV. A detailed calculation is shown
in the appendix.

\section{Effective Lagrangian}
\label{sec:anomalous} New physics effects on phenomena at the
electroweak scale can be described by the higher dimensional
operators~\cite{buchmuller}. The effective Lagrangian is given by
\begin{align}
{\mathcal L}_\mathrm{eff} &= {\mathcal
L}_\mathrm{SM}+\sum_i\sum_{n\ge 5}
\frac{C_i}{\Lambda^{n-4}}{\mathcal O}_i^{(n)},
\end{align}
where ${\mathcal L}_\mathrm{SM}$ is Lagrangian of the SM, $C_i$
are the coupling strengths of the dimension-$n$ operators
${\mathcal O}_i^{(n)}$, and $\Lambda$ is a cut off scale of the
SM. The coefficients of these higher dimensional operators can in
principle be calculated by assuming new physics models which are
defined above the scale $\Lambda$. When $\Lambda$ is much greater
than the electroweak scale, the dimension-six operators can give
leading contributions to the deviations from the SM.

If the top-Yukawa interaction is modified by the dimension-six
operators, its effect can be observed in the processes of $gg\to H$
and $gg\to HH$.
The dimension-six operators relevant to the gluon fusion mechanism are
\begin{align}
{\mathcal O}_{t1} &=
\left(\Phi^\dag\Phi-\frac{v^2}2\right)\left({\overline
Q}\,t_R^{}\,
\widetilde{\Phi}+\widetilde{\Phi}\,\overline{t_R^{}}\,Q\right),\\
{\mathcal O}_{Dt} &=
\left(\overline{Q}\,D_\mu\,t_R^{}\right)D^\mu\widetilde{\Phi}
+\left(D^\mu\widetilde{\Phi}\right)^\dag\left(\overline{D_\mu\,t_R^{}}\,
Q\right),\\
{\mathcal O}_{tG\Phi} &=
\left[\left(\overline{Q}\,\sigma^{\mu\nu}\lambda^A\,t_R^{}\right)
\widetilde{\Phi}+\widetilde{\Phi}^\dag\left(\overline{t_R^{}}\,
\sigma^{\mu\nu}\lambda^A\,Q\right)\right] G^A_{\mu\nu},
\end{align}
where $Q=(u,d)^T_L$, $G^A_{\mu\nu}$ is the field strength of
gluons with $SU(3)$ generators $\lambda^A(A=1$--$8)$, $\Phi$ is a
scalar-iso-doublet with hypercharge $Y=1/2$,
$\widetilde{\Phi}=i\,\tau_2\Phi^*$, and $v$ ($\sim 246$ GeV) is
the vacuum expectation value whose origin may come from color blind
dynamics. The effective top-Yukawa
coupling deviates from the SM value due to ${\mathcal O}_{t1}$
and ${\mathcal O}_{Dt}^{}$.
There is only one dimension-six operator ${\mathcal
O}_G^{}$~\cite{Pierce:2006dh} that contributes to $ggH$ and $ggHH$
vertices at the tree level; i.e.,
\begin{align}
{\mathcal O}_G^{} &= \left(\Phi^\dag\Phi-\frac{v^2}2\right)
G^A_{\mu\nu}{G^{A}}^{\mu\nu}, \label{Eq:OG}
\end{align}
whose origin can come from colored new dynamics at the TeV scale.

The triple-Higgs boson coupling also contributes to double-Higgs
production $gg\to HH$. Dimension-six genuine-Higgs operators
change the Higgs self-coupling, which are given
by~\cite{Barger:2003rs}
\begin{align}
{\mathcal O}_{\Phi1} &=
\frac12\,\partial_\mu\left(\Phi^\dag\Phi\right)
\partial^\mu\left(\Phi^\dag\Phi\right),\\
{\mathcal O}_{\Phi2}
&= -\frac13\left(\Phi^\dag\Phi\right)^3,\\
{\mathcal O}_{\Phi3} &=
\left(D_\mu\Phi\right)^\dag\Phi\,\Phi^\dag\left(D^\mu\Phi\right).
\end{align}
Normalization of the Higgs field is shifted by introduction of
the operator ${\mathcal O}_{\Phi1}$. All the Higgs interactions are
corrected after the wave function renormalization for the Higgs
boson~\cite{Barger:2003rs}.
The Higgs potential is modified by ${\mathcal O}_{\Phi2}$;
then the triple-Higgs boson coupling
is a function of $C_{\Phi2}^{}$ and the Higgs boson mass.
The coefficient $C_{\Phi2}^{}$ simply shifts the triple-Higgs
coupling.

The coefficients $C_{t1}$, $C_{\Phi1}$ and $C_{\Phi2}$ are free
from the current experimental data\footnote{The coefficient $C_{\Phi1}$
would be determined precisely at the ILC by using the gauge boson
association processes such as Higgs-strahlung and vector
boson fusion.}.
In contrast, the coefficient of ${\mathcal O}_{\Phi3}$ (${\mathcal
O}_{Dt}^{}$) contributes to the electroweak rho parameter at the tree level
(and the one-loop level), which is strongly constrained by the
experimental data~\cite{gounaris2,Kanemura:2006bh}. The effects of
${\mathcal O}_{tG\Phi}$ will be measured by top-pair production, 
$gg\to t{\bar t}$. Here we neglect the operators ${\mathcal O}_{Dt}^{}$,
${\mathcal O}_{tG\Phi}$ and ${\mathcal O}_{\Phi3}$ in the following
discussion assuming that these operators would be well constrained by the
other processes, and we concentrate on the effects due to the operators
${\mathcal O}_{t1}, {\mathcal O}_{\Phi1}, {\mathcal O}_{\Phi2}$ and
${\mathcal O}_{G}$. In Ref.~\cite{Pierce:2006dh}, the bounds on the
operator ${\mathcal O}_{G}$ are evaluated through the process
$gg\to H \to WW$ for $m_H^{}\gtrsim 160$ GeV,
\begin{align}
-1.2\lesssim a_G^{}\left(\frac{\alpha_s}{4\pi}\right)^{-1}\lesssim
0.5,
\end{align}
where the scaled couplings are defined as
$a_i=C_i\frac{v^2}{\Lambda^2}$.

The theoretical upper bounds on these operators from tree level
unitarity~\cite{pwu} have been discussed in the
literature~\cite{gounaris1,Barger:2003rs}. Such bounds on these less
constrained operators are given by
\begin{align}
a_{t1}\lesssim \frac{16\pi}{3\sqrt2}\frac{v}{\Lambda},\\
a_{\Phi1,\Phi2}^{}\lesssim 4\pi\frac{v^2}{\Lambda^2}.
\end{align}
If we consider the low energy cut off $\Lambda=1$--$3$ TeV, the
upper bound on $a_{t1}$ are $3.0$--$1.0$, respectively~\cite{Barger:2003rs}.
\section{Numerical evaluation of the Higgs production processes}
\label{sec:} 
\begin{figure}[tb]
\centering
\includegraphics[width=7.5cm]{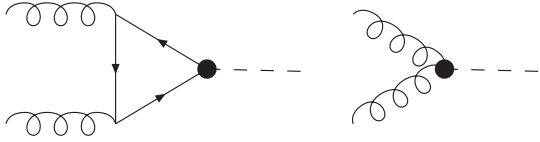}
\caption{Feynman diagrams for the single-Higgs production process
via gluon fusion. Curled, dashed and solid lines represent gluons,
Higgs bosons and quarks, respectively. Dots denote the new physics interaction.
In the SM, there is no tree level contact interaction.}
\label{FIG:FeynGGH}
\end{figure}

By introducing the dimension-six operators ${\mathcal
O}_{t1}$ and ${\mathcal O}_{\Phi1}$, the effective top-Yukawa
coupling is expressed as
\begin{align}
y_t^\text{eff}&=Z_{\Phi1}\left(\frac{\sqrt2m_t}{v}-a_{t1}\right),
\label{Eq:yteff}
\end{align}
where $Z_{\Phi1}=(1+a_{\Phi1})^{-1/2}$. Feynman diagrams for
single-Higgs production via gluon fusion is depicted in
FIG.~\ref{FIG:FeynGGH}.
For $a_G^{}=0$, the new physics contribution to this process
only appears in the effective top-Yukawa coupling.
Therefore $gg\to H$ only depends on $a_{t1}^{}$ and $a_{\Phi1}^{}$
by the combination given in Eq.~\eqref{Eq:yteff}.
We here define the parameter sets; Set A--Set F, for the coefficients
of dimension-six operators in TABLE~\ref{Tab:Sets}. Set A corresponds
to the SM. The effects of ${\mathcal O}_{t1}$ are studied by Set B and Set C,
and those of ${\mathcal O}_{t1}$ are investigated by Set D and Set E.
Set $F$ shows those of ${\mathcal O}_{\Phi2}$. The values of
$a_{t1}=\pm0.5$ correspond to the unitarity bounds for $\Lambda=5$
TeV, which change the effective top-Yukawa coupling by about $50$\%.
If we take a lower cut off scale ($\Lambda<5$ TeV), the unitarity
constraint on $a_{t1}$ becomes milder.
\begin{table}
\centering
\begin{tabular}{|c|c|c|c|c|c|c|}
\hline
Set&A&B&C&D&E&F\\\hline
$a_{t1}$&$0$&$+0.5$&$-0.5$&$0$&$0$&$0$\\\hline
$a_G^{}$&$0$&$0$&$0$&$+0.004$&$-0.004$&$0$\\\hline
$a_{\Phi2}^{}$&$0$&$0$&$0$&$0$&$0$&$+0.5$\\\hline
\end{tabular}
\caption{Parameter sets for the coefficients of the dimension-six operators.}
\label{Tab:Sets}
\end{table}

\begin{figure}[tb]
\centering
\includegraphics[width=7cm]{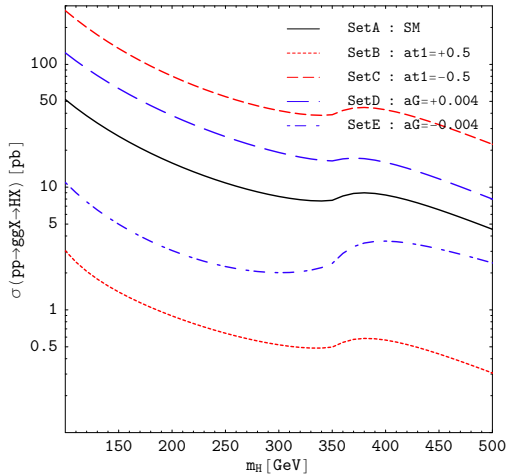}
\caption{The cross section of $pp\to ggX\to HX$
with $\sqrt{s}=14$ TeV as a function of the Higgs boson
mass. Curves denote the cross sections derived in the SM (Set A),
and in the SM with anomalous dimension-six couplings (Set B--Set E).}
\label{FIG:ggH}
\end{figure}

The hadronic production cross section for $pp\to ggX\to HX$ is evaluated
as a function of $m_H^{}$ at the LHC in FIG.~\ref{FIG:ggH}.
Detailed calculations are shown in Appendix A.
The solid, dotted, dashed, long-dashed and dot-dashed curves
correspond to the parameter sets Set A--Set E, respectively.
The peak around $m_H^{}\sim350$ GeV in these curves is understood as
the threshold effect due to the top-quark loop. If the
effective top-Yukawa coupling deviates from its SM value
(Set B and Set C), the cross section can be enhanced by a
factor $\sim9/4$ or suppressed by $\sim1/4$ for entire range of
the Higgs boson mass. These effects are determined only by
$y_t^\text{eff}$.
The differences from the SM for Set D and Set E are comparable to
that for top-Higgs coupling for Set B and Set C around $m_H^{}\sim
120$ GeV. The effects on the cross section from Set D and Set E are relatively
small compared to Set B and Set C for the larger Higgs boson masses.
These structures are realized by the interference of the amplitudes between
the new physics contributions and the SM one.

\begin{figure}[tb]
\centering
\begin{minipage}{0.49\hsize}
\includegraphics[width=7cm]{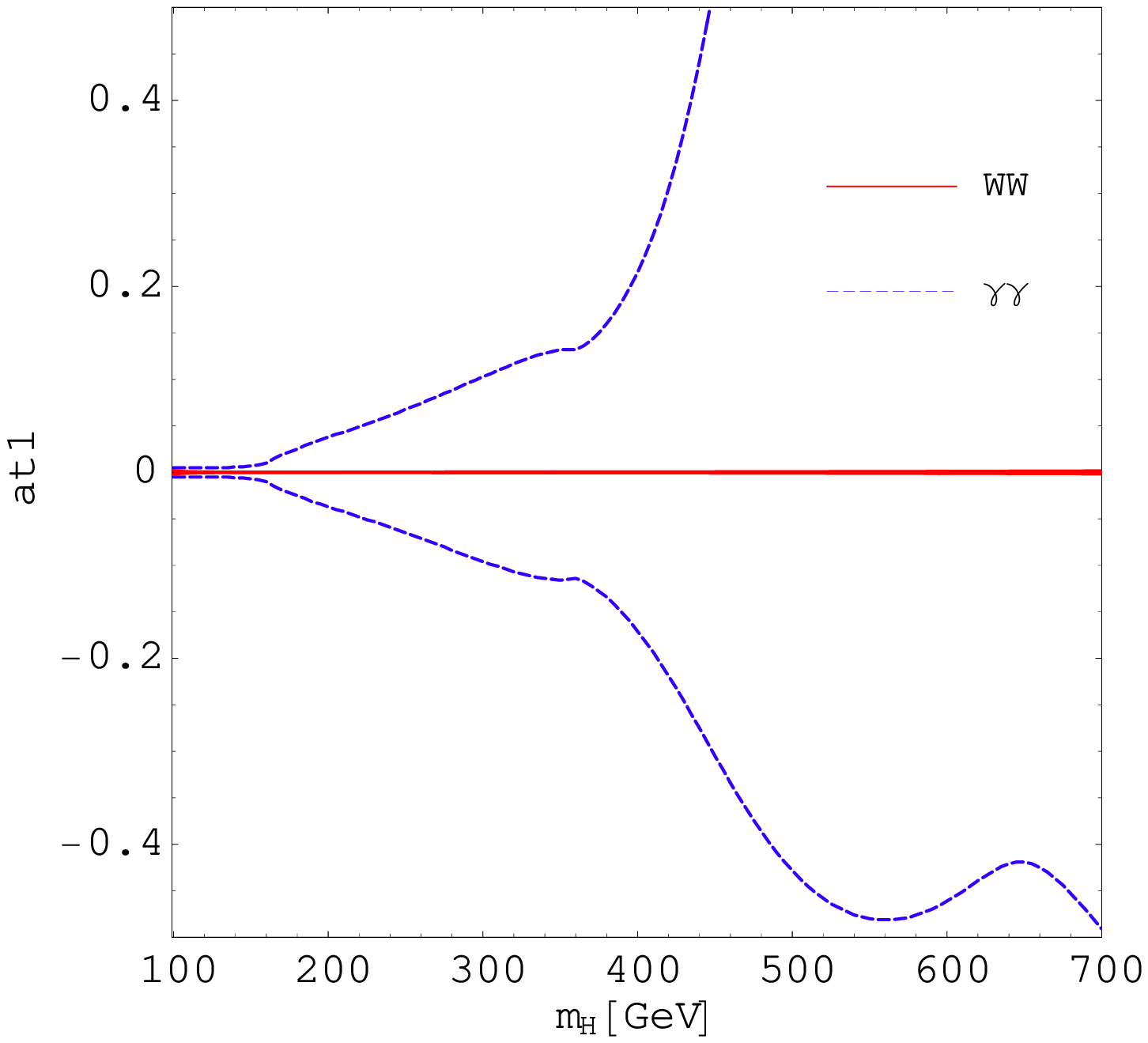}
\end{minipage}
\begin{minipage}{0.49\hsize}
\includegraphics[width=7cm]{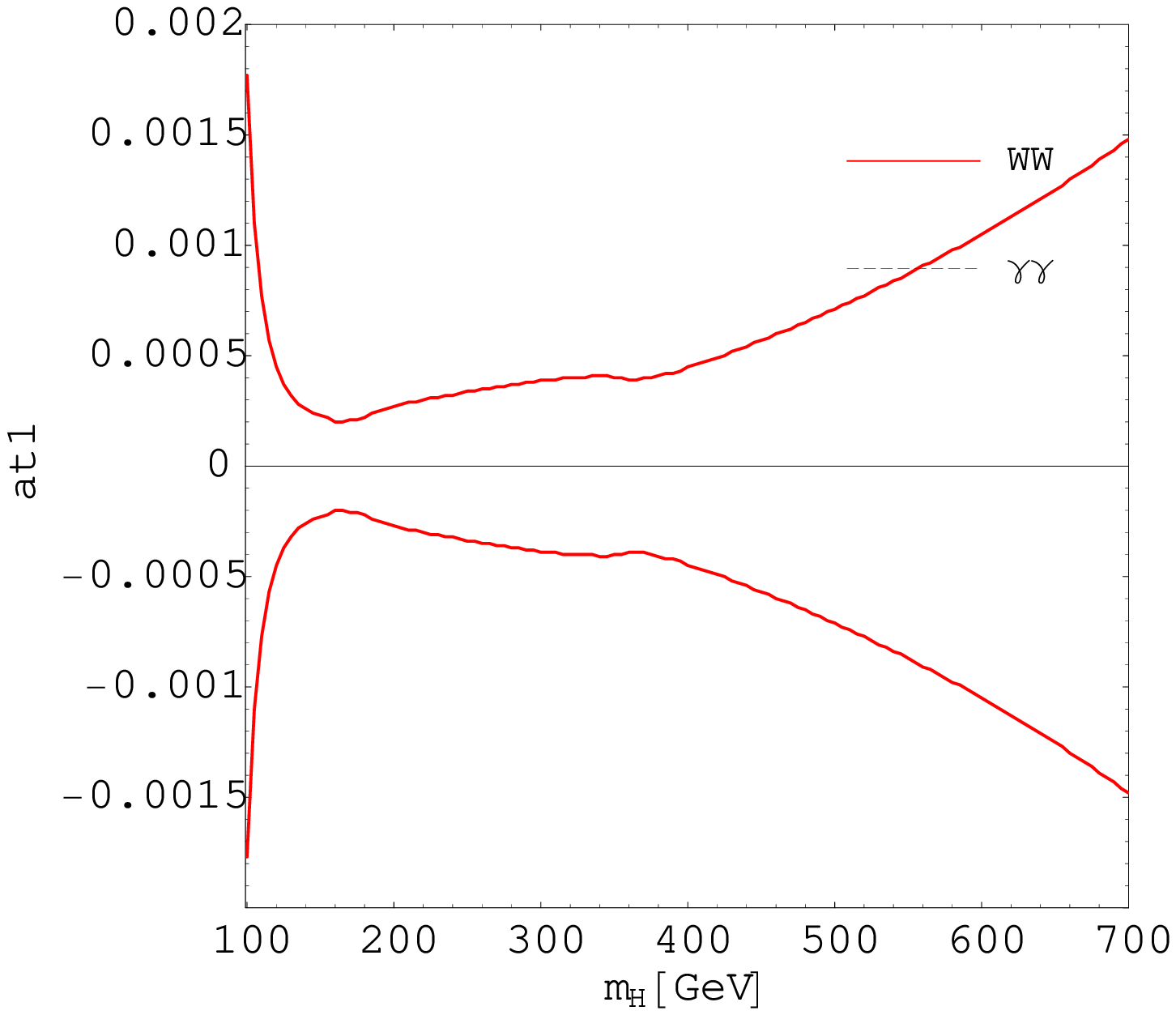}
\end{minipage}
\caption{The plot of the statistical sensitivity for $a_{t1}^{}$
on $N=L\sigma(pp\to ggX\to HX){\mathcal B}(H\to WW, \gamma\gamma)$
where the integrated luminosity is $L=300 \text{fb}^{-1}$.
Each curve denotes the $1\sigma$ deviation from the SM predictions.}
\label{FIG:GGH.sens.t1}
\end{figure}

\begin{figure}[tb]
\centering
\begin{minipage}{0.49\hsize}
\includegraphics[width=7cm]{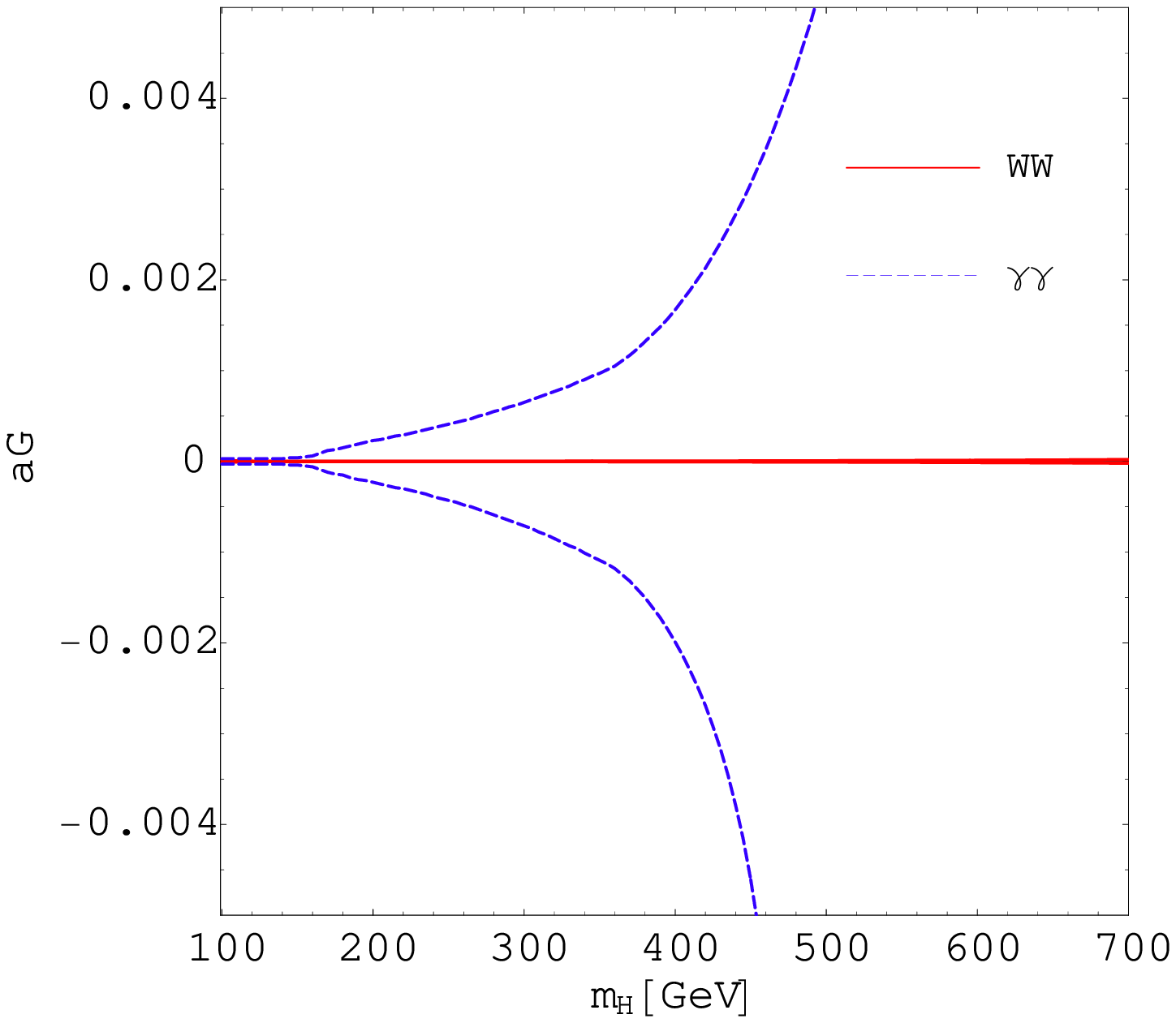}
\end{minipage}
\begin{minipage}{0.49\hsize}
\includegraphics[width=7cm]{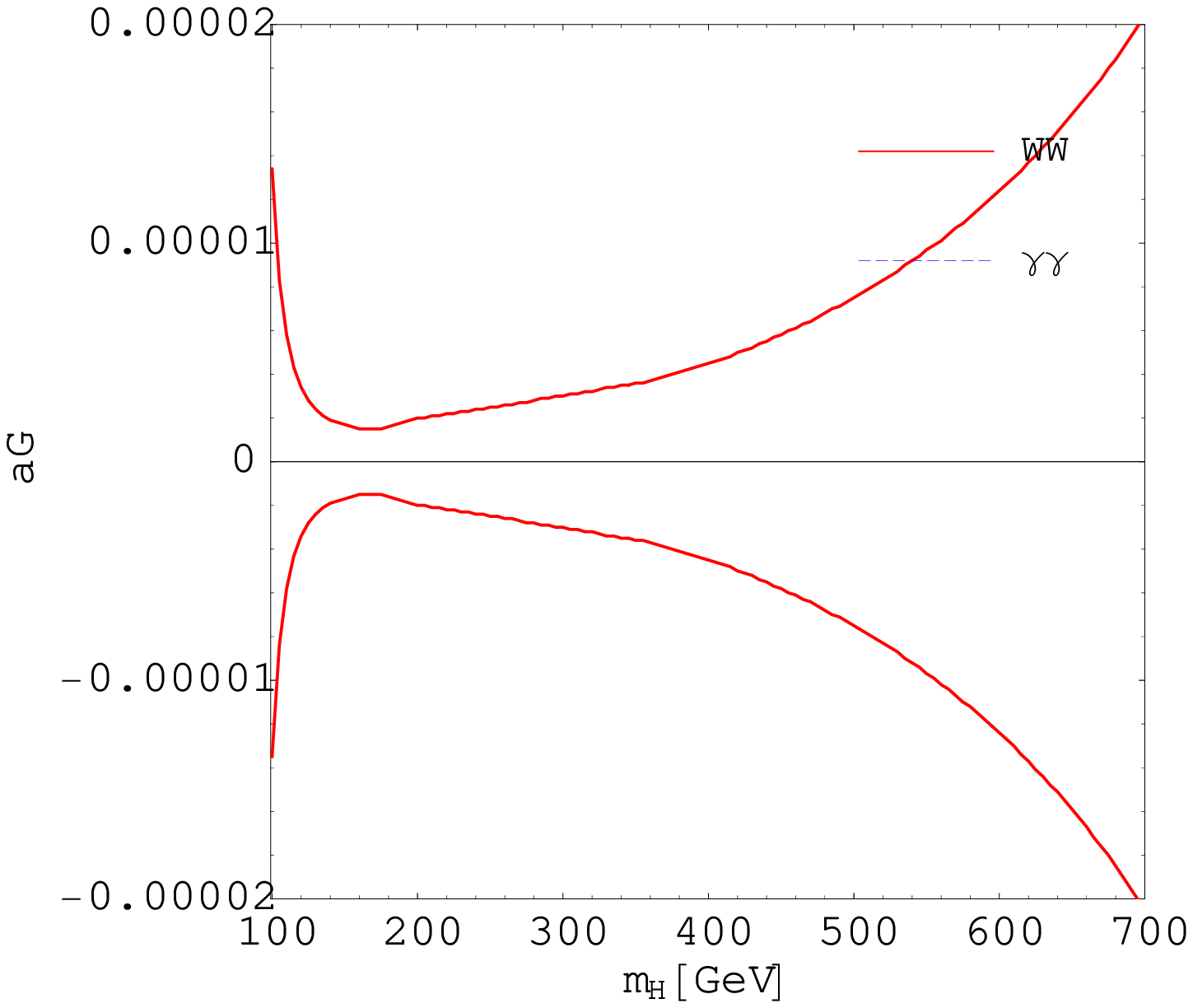}
\end{minipage}
\caption{The plot of the statistical sensitivity for $a_G^{}$
on $N=L\,\sigma(pp\to ggX\to HX){\mathcal B}(H\to WW, \gamma\gamma)$
where the integrated luminosity is $L=300 \text{fb}^{-1}$.
Each curve denotes the $1\sigma$ deviation from the SM predictions.}
\label{FIG:GGH.sens.G}
\end{figure}

In FIGs.~\ref{FIG:GGH.sens.t1} and \ref{FIG:GGH.sens.G}, we evaluate the
statistical sensitivities for anomalous parameters
on $N = L\,\sigma(pp\to ggX\to HX){\mathcal B}(H\to WW, \gamma\gamma)$
where the integrated luminosity is assumed to be $L=300 \text{fb}^{-1}$.
The efficiencies of $W$ bosons and photons are taken as $100\%$ for the illustration.
Therefore, in these plots, the $H\to WW$ decay mode always gives a better sensitivity
than the $H\to \gamma\gamma$ decay mode because the decay branching ratios hold
the relation ${\mathcal B}(H\to WW) \gg {\mathcal B} (H\to \gamma\gamma)$
for $m_H^{}\gtrsim 100$ GeV. Note that we do not include any backgrounds
to calculate the statistical sensitivity. For larger Higgs boson masses,
the sensitivities become worse due to the decreasing of the cross section.
For the $WW$ decay mode with $m_H^{}\lesssim 120$ GeV, it has also bad sensitivities
because of the small branching fraction of $H\to WW$.

\begin{figure}[tb]
\centering
\begin{minipage}{0.49\hsize}
\includegraphics[width=7cm]{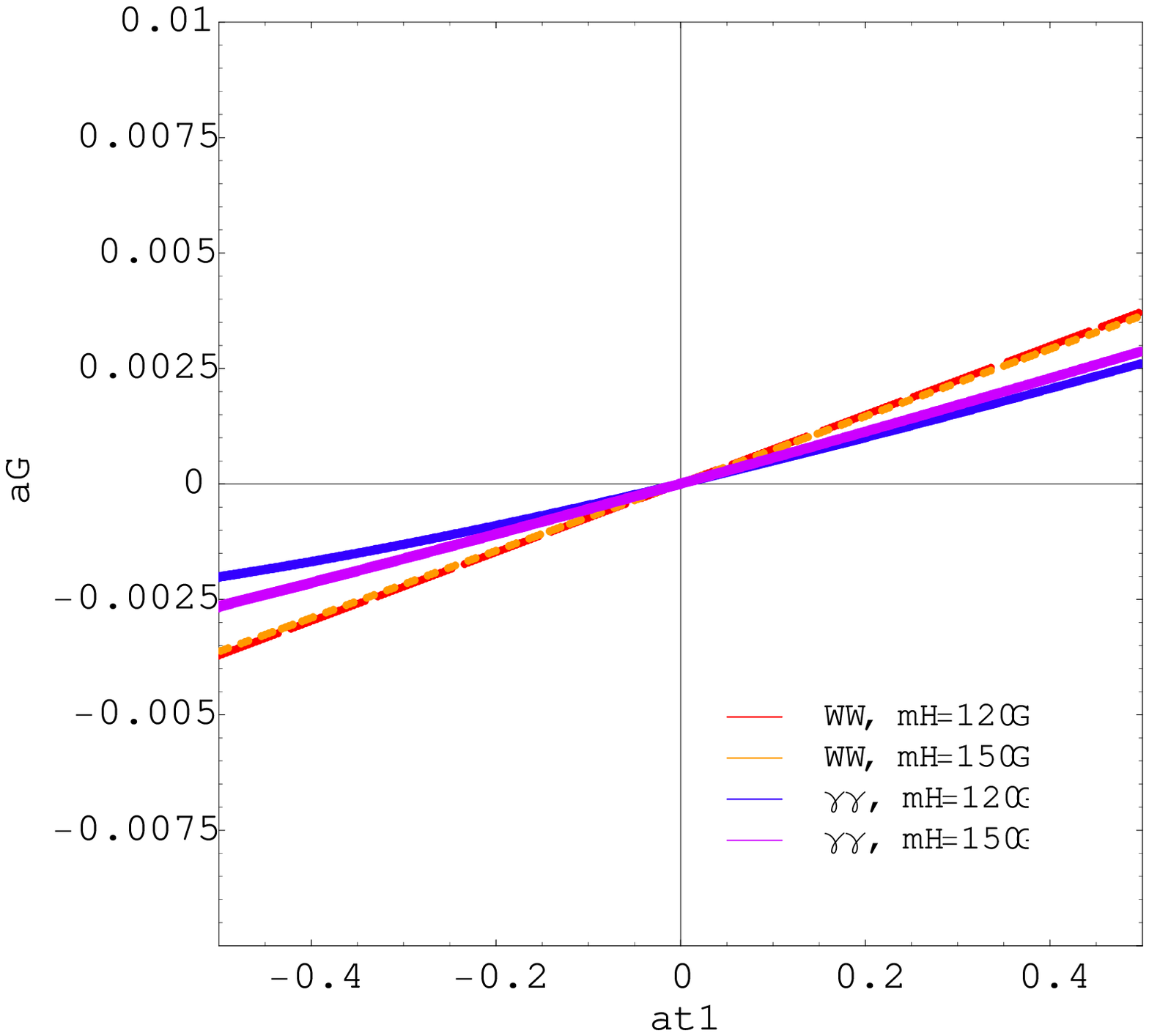}
\end{minipage}
\begin{minipage}{0.49\hsize}
\includegraphics[width=7cm]{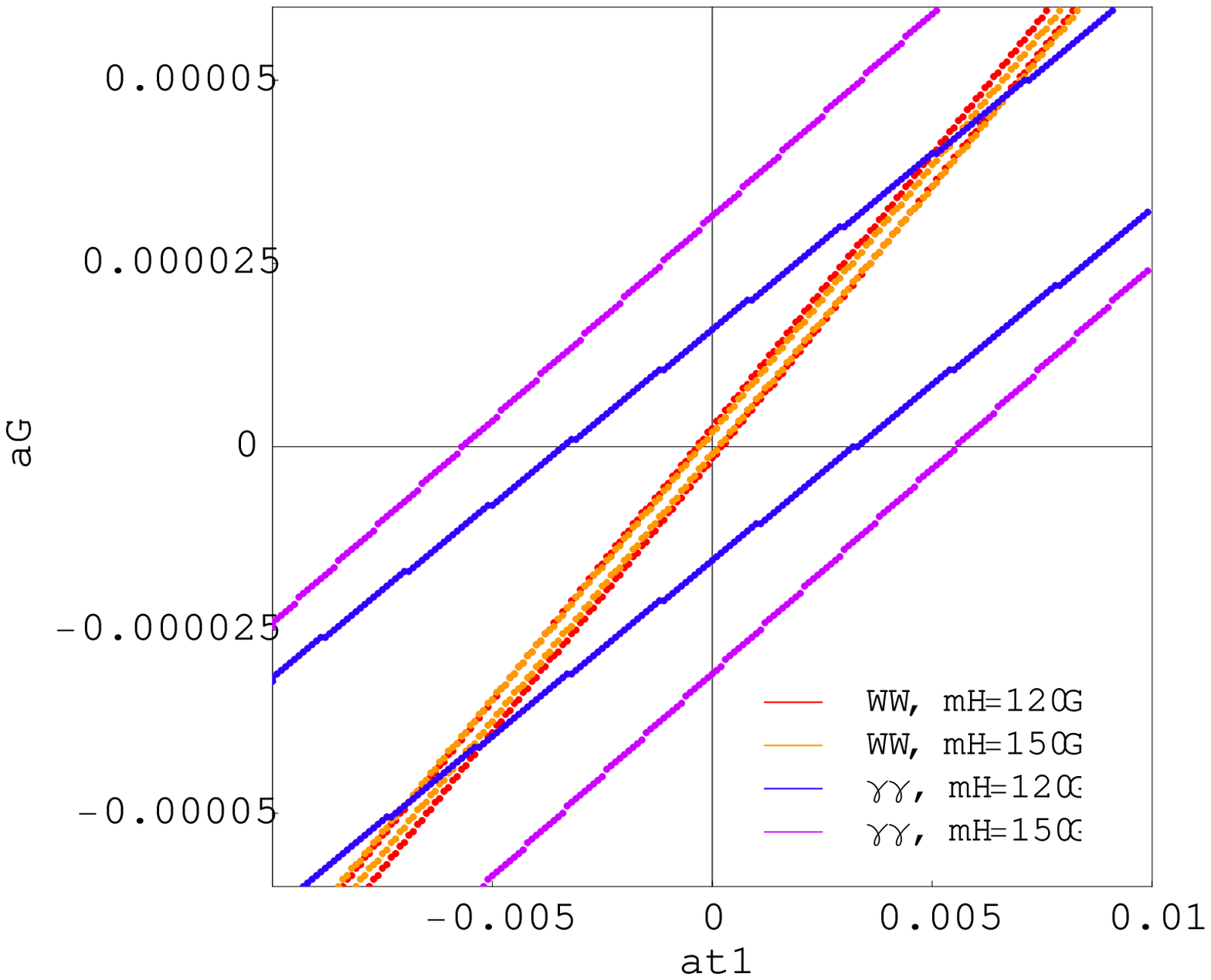}
\end{minipage}
\caption{The sensitivity plots in the $a_{t1}^{}$--$a_G^{}$ plane
on $N=L\,\sigma(pp\to ggX\to HX){\mathcal B}(H\to WW, \gamma\gamma)$
where the integrated luminosity is $L=300 \text{fb}^{-1}$.
Each contour represents the $1\sigma$ deviation from the SM predictions.}
\label{FIG:GGH.sens}
\end{figure}

In FIG.~\ref{FIG:GGH.sens}, we show the contour plot of the sensitivities in the
$a_{t1}^{}$--$a_G^{}$ plane. There is a strong correlation between $a_{t1}^{}$
and $a_G^{}$. This means that $a_{t1}^{}$ can mimic the effect of $a_G^{}$
in this process. It is understood by the destructive interference of
the top-loop diagram, which is shifted by $a_{t1}^{}$ and
the {\em tree level} diagram, which is induced by the dimension-six
operator ${\mathcal O}_G^{}$. Therefore, if the deviation from the SM
is found in $gg\to H$, we cannot distinguish the effects of
these anomalous couplings.

\begin{figure}[tb]
\includegraphics[width=15cm]{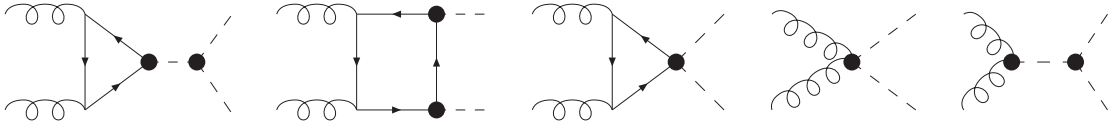}
\caption{Feynman diagrams for the double-Higgs production process
$gg\to HH$ are depicted. Dots represent the new vertices of the
dimension-six operators.}\label{FIG:FeynGGHH}
\end{figure}

Next, let us discuss the double-Higgs production.
In FIG.~\ref{FIG:FeynGGHH}, we show the Feynman diagrams for the
process $gg\to HH$.
Invariant amplitudes for the sub-process are given in Appendix A.
For each diagram, appropriate crossing of external Higgs-boson and
gluon lines should be taken into account.
In the SM, there are only two kinds of topology,
the first and second diagrams from the left.
This process would be used to measure the triple-Higgs boson
coupling at the SLHC. The vertex function of $HHH$ is modified
by the introduction of the genuine Higgs operators as
\begin{align}
\lambda_{HHH}^{}({\hat s}) =
Z_{\Phi1}^{3/2}\left(Z_{\Phi1}^{-1}\frac{m_H^2}{2v^2} -\frac{{\hat
s}+2m_H^2}{v^2}a_{\Phi1}+\frac{a_{\Phi2}}3\right)v, \label{Eq:HHH}
\end{align}
where $\sqrt{\hat s}$ is the center of mass energy of $gg\to HH$.
For Set F, the coefficient $a_{\Phi 2}^{}$ is taken to be positive
to ensure vacuum stability. The value $a_{\Phi 2}^{}=+0.5$
corresponds to the $140$\% enhancement of the triple-Higgs boson
coupling for $m_H^{}=120$ GeV.

\begin{figure}[tb]
\includegraphics[width=7cm]{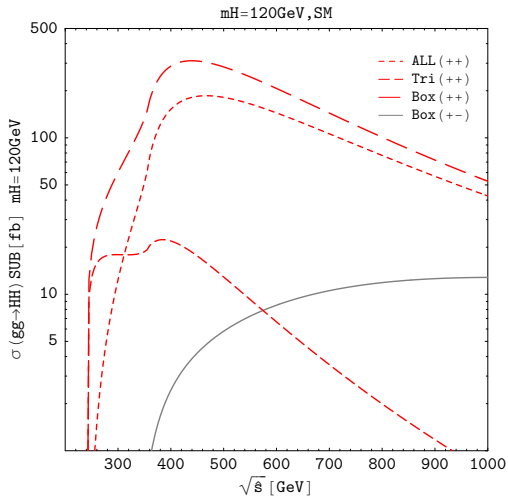}
\caption{The cross section for the Higgs pair production sub-process $gg\to HH$ as a function of scattering energy for
$m_H^{}=120$ GeV in the SM. The dashed (long-dashed) curve denotes
the contribution coming from the triangle (box) diagram with
the $(+,+)$ helicity set of gluons.
The dotted curve is a composition of these two.
The thin solid one represents the helicity set $(+,-)$ in the
box diagram.}\label{FIG:ggHH-sub-SM}
\end{figure}

The cross sections for the Higgs boson pair production via the
gluon fusion sub-process are shown in FIG.~\ref{FIG:ggHH-sub-SM}
in the SM.
The dotted and thin solid curves represent the cross section with the
helicity set $(+,+)$ and $(+,-)$ of the gluons. For $m_H^{}=120$ GeV,
the main contribution comes from the box diagram (long-dashed)
because the triangle diagram (dashed) contains the Higgs boson self-coupling,
which is proportional to the Higgs boson mass squared, and 
it has the typical behavior
of the $s$-channel process.

\begin{figure}[tb]
\begin{minipage}{0.49\hsize}
\includegraphics[width=7cm]{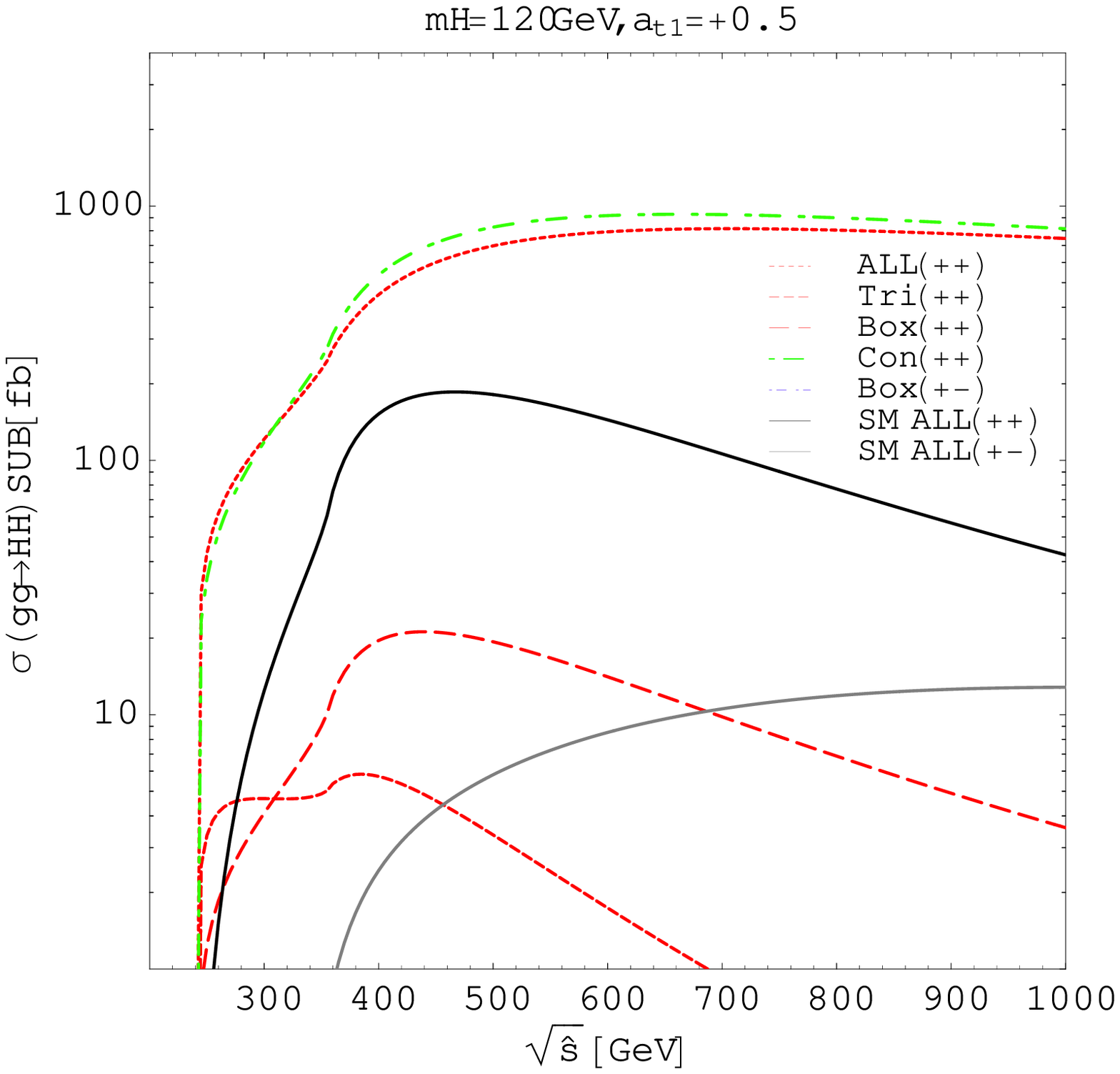}
\end{minipage}
\begin{minipage}{0.49\hsize}
\includegraphics[width=7cm]{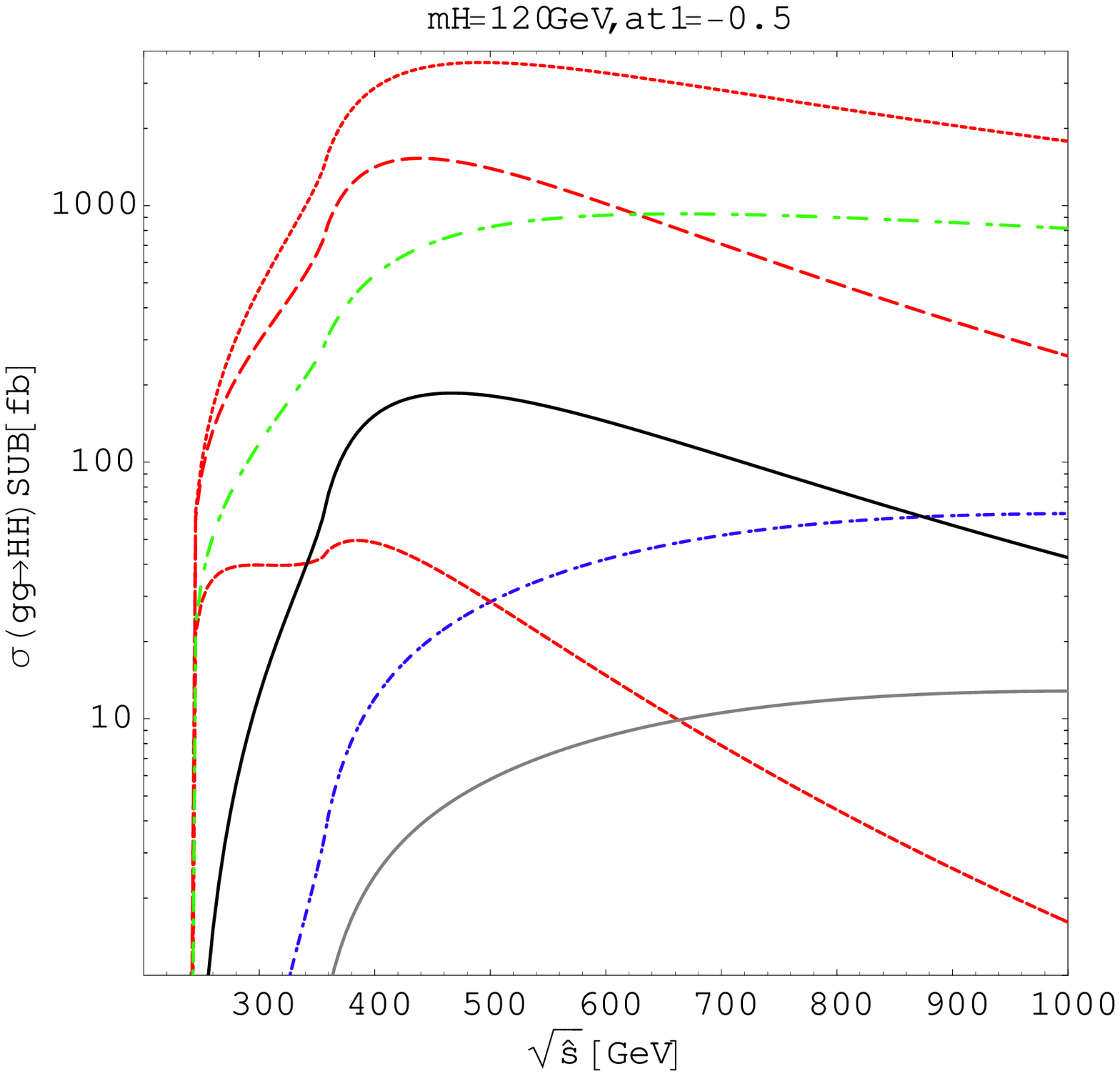}
\end{minipage}
\caption{The cross section for the process $gg\to HH$ as a function of
$\sqrt{\hat s}$ for $m_H^{}=120$ GeV for Set B and C.
Each curve is given in the
same manner as in FIG.~\ref{FIG:ggHH-sub-SM}. The long-dot-dashed curve
denotes the contribution comes from the new vertex which is induced by
the dimension-six operator. The SM with helicity sets $(+,+)$ and $(+,-)$
of the gluons are also shown in thick and thin solid curves as a
reference.}\label{FIG:ggHH-sub-t1}
\end{figure}

In FIG.~\ref{FIG:ggHH-sub-t1}, we show the effect of the
dimension-six top--Higgs interaction on $gg\to HH$ as a function of
$m_H^{}$ for Set B and Set C.
These curves are used as the same manner as
in FIG.~\ref{FIG:ggHH-sub-t1}. The additional long-dot-dashed
curve represents new vertex contribution. The SM with gluon
helicity sets $(+,+)$ and $(+,-)$ are also shown in thick and thin
solid curves. For Set B, it can be seen that the effective
top-Yukawa coupling is suppressed in both the triangle and the box
diagrams. The cross section can also be enhanced by the dimension-five
interaction vertex of $t{\bar t}HH$ due to ${\mathcal O}_{t1}$.
On the contrary, for Set C the
contributions from the box diagrams become large because of the
large effective top-Yukawa coupling. The synergy effect of both
the contributions enhances the sub-process cross section significantly.

\begin{figure}[tb]
\centering
\includegraphics[width=7cm]{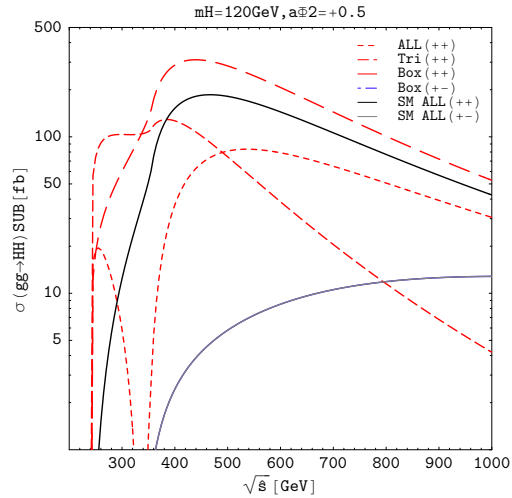}
\caption{The cross section of $gg\to HH$ for
$m_H^{}=120$ GeV with $a_{\Phi2}=+0.5$ as a function of
$\sqrt{\hat s}$. The curves are defined in the same
manner as FIG.~\ref{FIG:ggHH-sub-t1}.} \label{FIG:ggHH-sub-H}
\end{figure}

In FIG.~\ref{FIG:ggHH-sub-H}, we show the $gg\to HH$ cross section
as a function of $\sqrt{\hat s}$ for Set F.
Each curve is given as in FIG.~\ref{FIG:ggHH-sub-t1}.
The contribution of $a_{\Phi2}^{}$ only appears in the triple-Higgs 
boson vertex. Around $m_H^{}\sim 350$ GeV, a strong destructive
interference between the triangle and the box diagrams occurs. The
effect in Set F is relatively small as compared to those in
Set B and Set C. There is also enhancement on the sub-process cross
section near the threshold of $HH$ production. These effects turn out to
give larger contributions to the hadronic cross section.

\begin{figure}[tb]
\includegraphics[width=7cm]{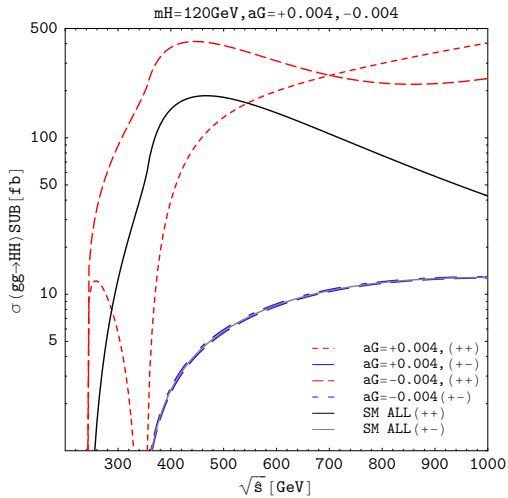}
\caption{The cross section of $gg\to HH$ as a function of sub-process 
energy for $m_H^{}=120$ GeV in the SM with
$a_G^{}=\pm0.004$. The dotted and long-dashed curves denote the gluon
helicity sets $(+,+)$ and $(+,-)$ for Set D. The dashed and dot-dashed
curves are those for Set E. The SM predictions are also given by solid curves.}\label{FIG:ggHH-sub-G}
\end{figure}

For completeness, we also show the case for Set D and Set E in
FIG.~\ref{FIG:ggHH-sub-G}.
The dotted and long-dashed curves represent the helicity
set $(+,+)$ and $(+,-)$ of gluons for Set D. Those for Set E are given by
the dashed and dot-dashed curves. The SM prediction is also
shown. For Set D, there is a cancellation between the
SM contribution and the anomalous tree level vertex $ggH$ in the $t{\bar
t}$ threshold region. In Set E, each contribution
is constructive in the same parameter region.

Convoluting the CTEQ6M parton distribution
function~\cite{cteq6}, the full cross section is evaluated
 for double-Higgs production
via gluon fusion.
\begin{figure}[tb]
\begin{minipage}{0.49\hsize}
\includegraphics[width=7cm]{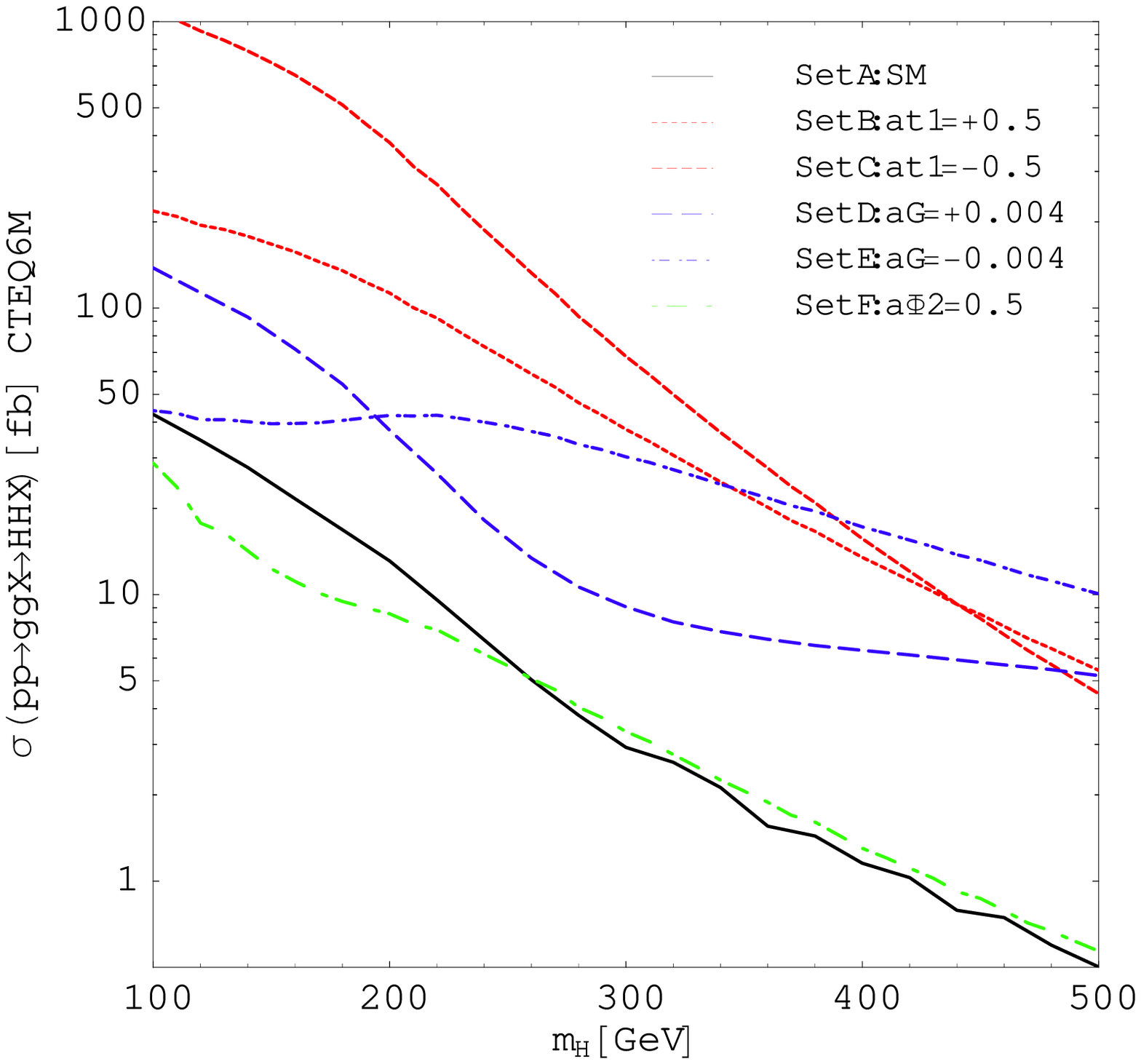}
\end{minipage}
\begin{minipage}{0.49\hsize}
\includegraphics[width=7cm]{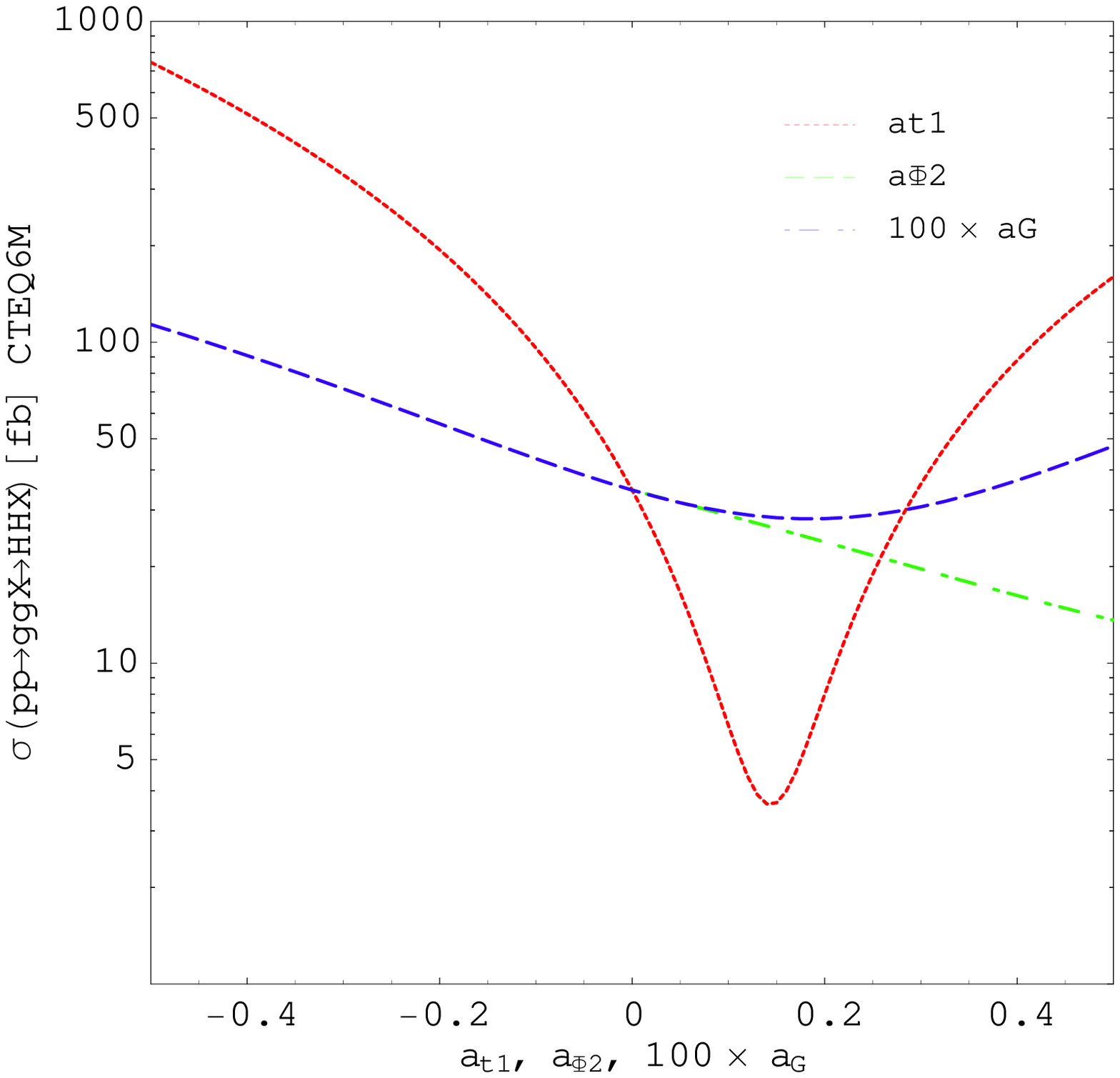}
\end{minipage}
\caption{The total cross section of the double-Higgs production
$pp\to ggX\to HHX$ as a function of the Higgs boson mass in the left figure and the
anomalous couplings in the right figure with center of
mass energy $\sqrt{s}=14$ TeV.} \label{FIG:ggHH-full}
\end{figure}
In FIG.~\ref{FIG:ggHH-full}, we show the hadronic cross sections
as a function of the Higgs boson mass in the left figure and as a function of
the dimension-six couplings in the right figure.
Significant enhancement from the dimension-six
operators ${\mathcal O}_{t1}$ and ${\mathcal O}_{G}^{}$ can be seen
in both figures for a wide range of parameter space. For Set F,
the curve and that of the SM one coincide for large $m_H^{}$ values
whose structure is easily understood from Eq.~\eqref{Eq:HHH}.
The effect of $a_{\Phi2}$ is reduced when we take larger
Higgs boson masses. The contributions from $a_{\Phi1}$ might be much
larger than those of the other operators. This effect will be
first examined by gauge boson association processes.

\begin{figure}[tb]
\centering
\begin{minipage}{0.49\hsize}
\includegraphics[width=7cm]{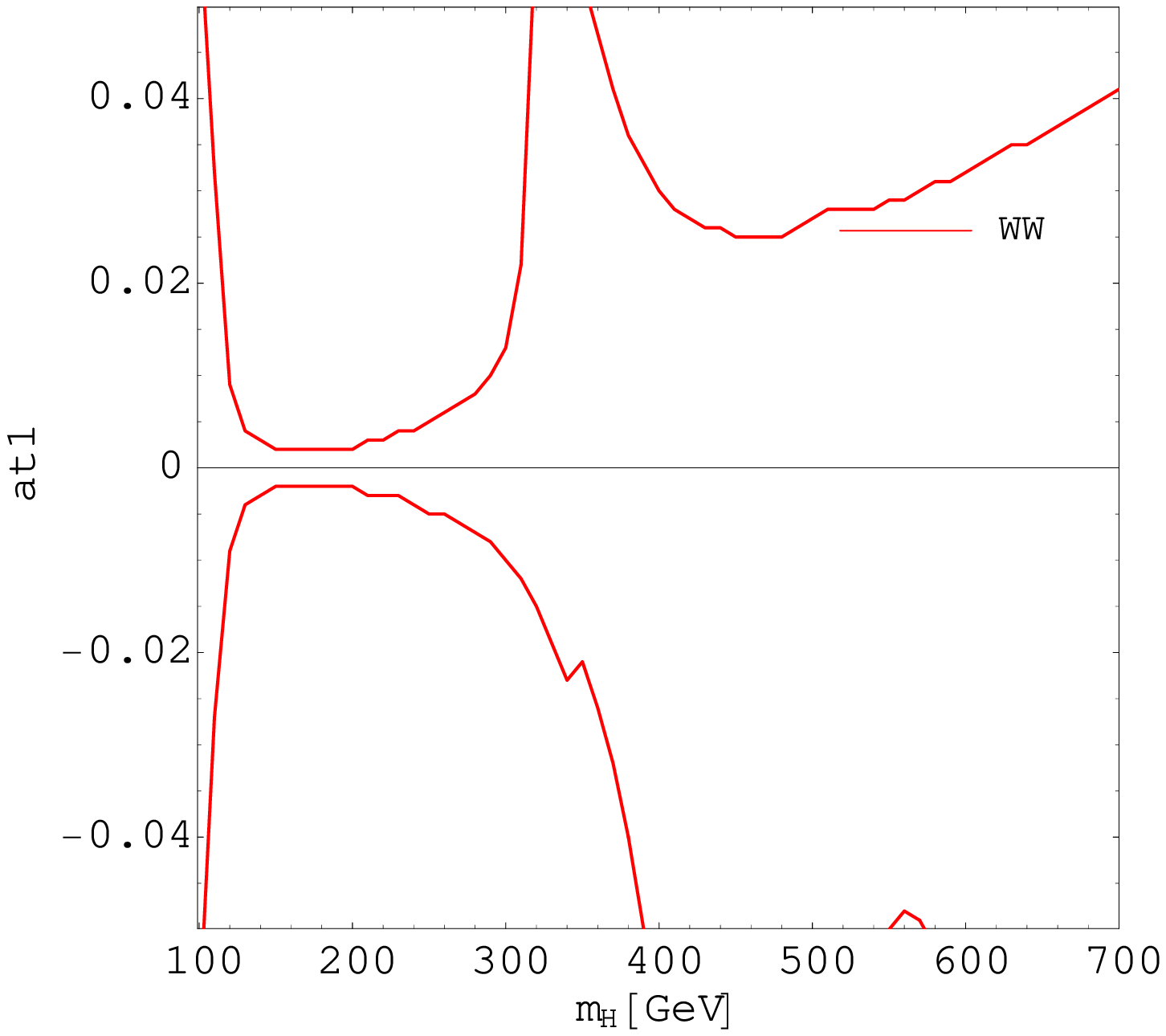}
\end{minipage}
\begin{minipage}{0.49\hsize}
\includegraphics[width=7cm]{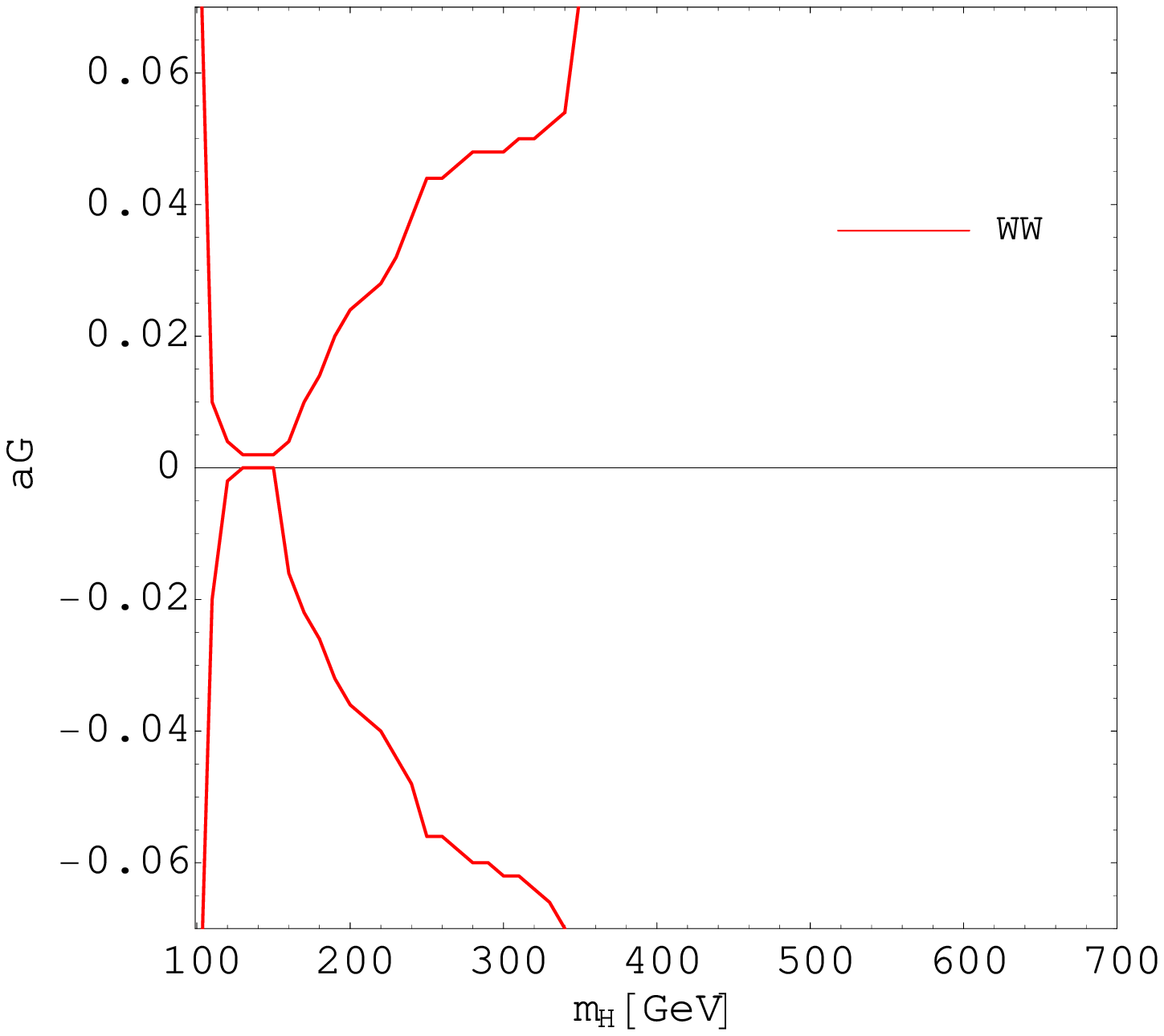}
\end{minipage}
\caption{The plot of the statistical sensitivity for $a_{t1}^{}$
on $N=L\sigma(pp\to ggX\to HHX){\mathcal B}(H\to WW){\mathcal B}(H\to WW)$
where the integrated luminosity is $L=300 \text{fb}^{-1}$.
Each curve denotes the $1\sigma$ deviation from the SM predictions.}
\label{FIG:GGHH.sens}
\end{figure}

In FIG.~\ref{FIG:GGHH.sens}, we evaluate the statistical sensitivities
for the anomalous parameters on
$N = L\,\sigma(pp\to ggX\to HHX){\mathcal B}(H\to WW){\mathcal B}(H\to WW)$
where the integrated luminosity is assumed to be $L=300 \text{fb}^{-1}$.
We focus on the anomalous parameters $a_{t1}^{}$ and $a_G^{}$,
because this process is insensitive to $a_{\Phi 2}^{}$ as we showed
in FIG.~\ref{FIG:ggHH-full}.
Obtained sensitivities are less smaller than those in $gg\to H$ process.
However, this process is still sensitive to $a_{t1}^{}$ for $m_H^{}\sim 150$ GeV
on some level due to the large enhancement (suppression) in the box diagram.
On the other hand, it is insensitive to the anomalous parameter $a_G^{}$.

\begin{figure}[tb]
\centering
\includegraphics[width=7cm]{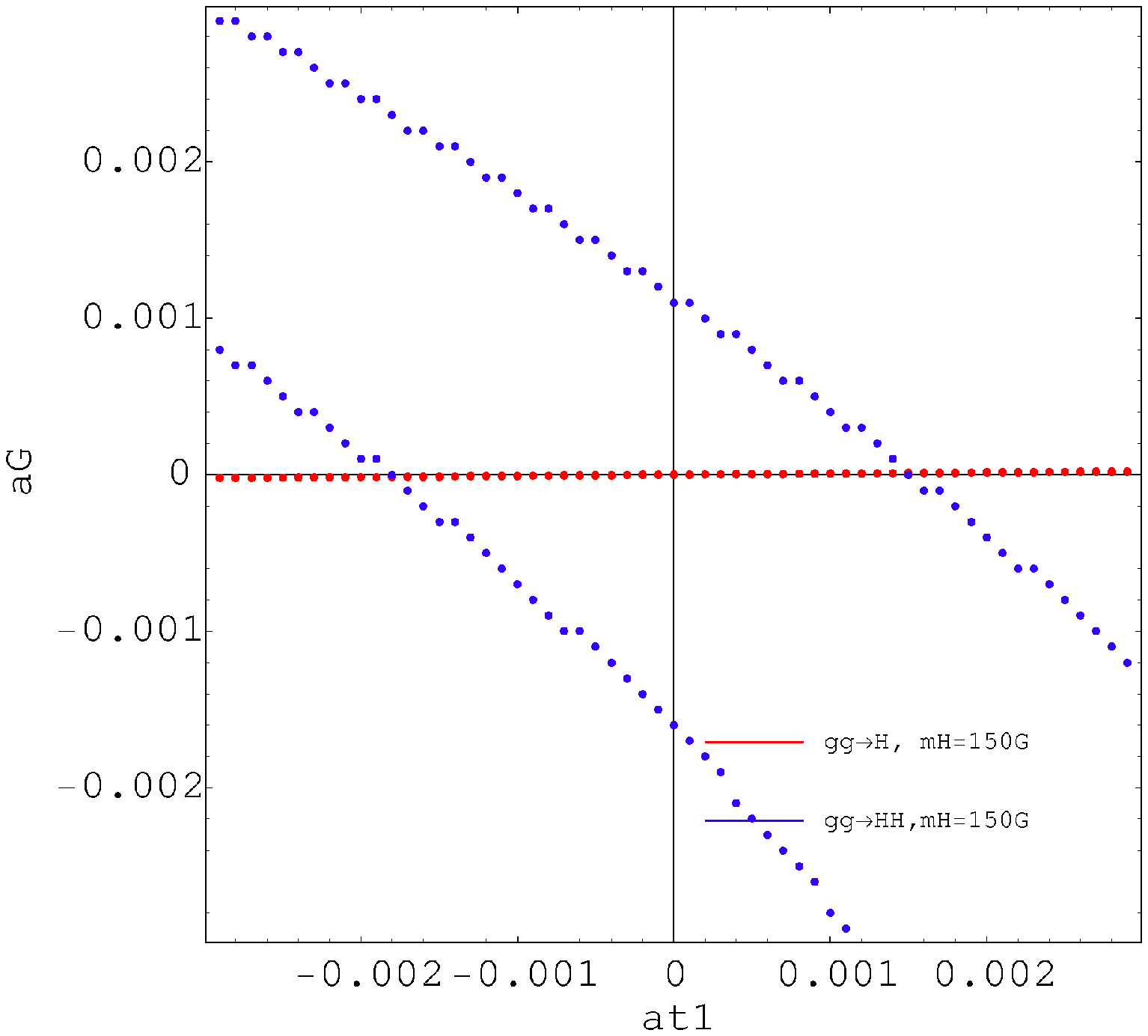}
\caption{The sensitivity plots in the $a_{t1}^{}$--$a_G^{}$ plane
on $N=L\,\sigma(pp\to ggX\to HHX){\mathcal B}(H\to WW){\mathcal B}(H\to WW)$
where the integrated luminosity is $L=300 \text{fb}^{-1}$ and $m_H^{}=150$ GeV.
Each contour represents the $1\sigma$ deviation from the SM predictions.
We also show the contour of the sensitivity on 
$N=L\,\sigma(pp\to ggX\to HX){\mathcal B}(H\to WW)$ } as a reference.
\label{FIG:GGHH.sens.con}
\end{figure}

We show the contour plot of the sensitivities in the $a_{t1}^{}$--$a_G^{}$ plane
in FIG.~\ref{FIG:GGHH.sens.con}.
To compare the single- and double-Higgs production processes,
we also show the results calculated from $gg\to H$.
By using the insensitivity of $gg\to HH$ to $a_G^{}$,
the anomalous parameter $a_{t1}^{}$ can be constrained.

The possibility of measuring the Higgs boson pair production has been
discussed in Ref.~\cite{Baur:2003gp} to determine the triple Higgs
boson coupling constant. Their background analyses can apply to our setup because
the dominant decay modes of the Higgs boson are almost the same as in
the SM. In our
case, the double-Higgs production process can be observable if the
cross section receives the above enhancement. However it does not
mean the improvement of the sensitivity for the triple-Higgs
boson coupling. The new vertex $t{\bar t}HH$ smears the effect of the Higgs
boson self-coupling. We find that $gg\to HH$ can still be sensitive to the
dimension-six top-Higgs interaction.
\section{Conclusions and discussions}
\label{sec:conclusion} In this paper, we have studied the impact
of the dimension-six operators on the Higgs production processes
via gluon fusion, i.e., $gg\to H$ and $gg\to HH$ at the LHC.
Constraints from the current experimental data and the theoretical
consistencies on the dimension-six operators are taken into
account.
We find that the contribution from the dimension-six top--Higgs
operators to single-Higgs production can significantly change the
cross section by a factor.
The double-Higgs production process can also receive a large
enhancement from the anomalous top-Higgs couplings.
The shift of the effective top-Yukawa coupling can
enhance the cross section significantly. In addition, the new
diagrams from the tree level vertex $t{\bar t}HH$ from ${\mathcal O}_{t1}^{}$
result in much larger cross sections than that in the SM.
Combined results of single- and double-Higgs production can
be used to discriminate between the
effects of dimension-six operators ${\mathcal O}_{t1}^{}$ and
${\mathcal O}_G^{}$.

Finally, we comment on the potential for the coupling measurements
at the international linear collider. As an optional process
the photon--photon collision $\gamma\gamma\to HH$ has a similar
structure to $gg\to HH$~\cite{Jikia}.
This process includes not only the top-quark loop but also
the $W$ boson loop, so that this kind of enhancement from
the top--Higgs interaction may be weakened by the $W$ boson loop. The
coupling $a_{\Phi2}^{}$ will also be measured at
double-Higgs-strahlung $e^-e^+\to ZHH$~\cite{Gounaris:1979px}, and W boson fusion
$e^-e^+\to\nu{\bar \nu}HH$~\cite{Kilian}, as well as above the photon--photon
collision~\cite{Cornet}.

\vspace{1cm} \noindent
{\large \it Acknowledgments}\\
S.K. was supported in part by Grants-in-Aid for Science
Research, Japan Society for the Promotion of Science No. 18034004.

\appendix
\section{Cross sections with dimension six operators}
We present formulae for the single- and the double-Higgs
production processes via the gluon fusion mechanism.

The hadronic cross section at the leading order is calculated
by convoluting with the parton distribution function (CTEQ6M) as
\begin{align}
\sigma^\text{LO}(gg\to H) \simeq
\frac{\pi^2}{8m_H^3}
\Gamma_{H\to gg}^{}\tau
\int_{\tau}^1\frac{dx}{x}g(x)g(\tau/x),
%
\end{align}
where $\tau=m_H^2/s$, and $g(x)$ is the gluon distribution function in a proton.
In the SM with higher dimensional operators,
the decay rate for the Higgs boson into gluons is calculated as
\begin{align}
&\Gamma_{H\to gg}^{} = \frac{G_Fm_H^3}{\sqrt2\pi}
\left|a_G^{}
+\frac{\alpha_s(\mu=m_H)}{8\pi}\left(1-\frac{a_{t1}v}{\sqrt2m_t}\right)
\frac{4m_t^2}{m_H^2}\left[2-m_H^2\left(1-\frac{4m_t^2}{m_H^2}\right)
C_0(m_H^2)\right]\right|^2,
\end{align}
where $C_0({\hat s})=C_0(0,0,{\hat s},m_t^2,m_t^2,m_t^2)$ and
$\alpha_s(\mu)$ is the running strong coupling constant.
We here use the Passarino--Veltman functions for
loop calculations~\cite{Passarino:1978jh}.
Large higher order corrections are known for
the gluon fusion mechanism~\cite{gghNLO,gghHO}.
The next to the leading order (NLO) correction for the cross section
can be treated by the K-factor~\cite{gghNLO},
\begin{align}
K^\text{NLO}_{gg\to H}
=\frac{\sigma^\text{NLO}(gg\to H)}{\sigma^\text{LO}(gg\to H)}
\simeq 1+\frac{\alpha_s(\mu)}{\pi}\left(\pi^2+\frac{11}2\right).
\end{align}
A naive introduction of dimension-six operator breaks
renormalizability of the theory. However, these operators can be
embedded in the more fundamental theory. Thus we here adopt
the correction only from gluon emissions to the effective $ggH$ vertex
in the heavy top-quark mass limit as the NLO correction.

In the effective theory, loop integrations by four momenta are
cut off at a some new physics scale $\Lambda$,
\begin{align}
S_n^\Lambda
&= \frac{(4\pi)^2}{i}\int\frac{d^4k}{(2\pi)^4}
\frac1{(k^2-C)^n}
=(-1)^nC^{2-n}\int_0^{\Lambda^2/C}dt\,\frac{t}{(1+t)^n}.
\end{align}
These integrals are related to those in the dimensional regularization (DR) as
\begin{align}
C_0^\Lambda
&= \int_0^1dx\int_0^{1-x}dy\left[-\frac1{H_C}
\frac1{\left(1+H_C/\Lambda^2\right)^2}\right]
\simeq C_0^\mathrm{DR}+\frac1{\Lambda^2},\\
D_0^\Lambda
&= \int_0^1dx\int_0^{1-x}dy\int_0^{1-x-y}dz
\left[\frac1{H_D^2}\frac{1+3H_D/\Lambda^2}{\left(1+H_D/\Lambda^2\right)^3}
\right] \simeq D_0^\text{DR}-\frac3{2\Lambda^2},
\end{align}
with
\begin{align}
&H^C
= (x\,r_1+y\,r_2)^2-x\,r_1^2-y\,r_2^2+m_t^2, \\
&H^D_{}
= (x\,r_1+y\,r_2+z\,r_3)^2-x\,r_1^2-y\,r_2^2-z\,r_3^2+m_t^2,
\end{align}
where $r_i=\sum_ip_i$. The $D_0$ function will appear in the
$gg\to HH$ cross section. We should comment on the reduction formulae of the
loop integrals in the Passarino--Veltman technique. These reduction formulae are
fully supported by Lorentz invariance, but the cut off regularization
generally violates it. In our analysis, we omit the effect of
the cut off $\Lambda$ in the loop integrals.
The corrections due to the cut off are small $(v/\Lambda)^2$
when the scale $\Lambda$ is set to be more than $3$ TeV,
so that the effects of the error turn out to be
numerically unimportant.

Let us discuss the helicity cross section for the sub-process $gg\to HH$.
The differential cross section is calculated as
\begin{align}
\frac{d{\hat \sigma}(g_{\lambda}g_{\lambda'}\to HH)}{d{\hat t}}
= \frac{\left|{\mathcal M}^{\lambda\lambda'}\right|^2}{16\pi{\hat s}^2}.
\end{align}
In the SM with dimension six operators ${\mathcal O}_{t1}, {\mathcal O}_{\Phi2}$
and ${\mathcal O}_{G}^{}$, helicity amplitudes ${\mathcal M}^{\lambda\lambda'}$
are given by
\begin{align}
{\mathcal M}^{++}&={\mathcal M}^{--}\nonumber \\
&= \frac{\alpha_sm_tv}{\sqrt2\pi}\left[2-{\hat s}\left(1-\frac{4m_t^2}{m_H^2}\right)C_0({\hat s})\right]
\left\{\frac{3a_{t1}}{v^2}+\frac{6(\sqrt2m_t/v-a_{t1})}{{\hat s}-m_H^2}
\left[\frac{m_H^2}{2v^2}+\frac{a_{\Phi2}^{}}3\right]\right\}
\nonumber \\
&\quad+\frac{\alpha_s(\sqrt2m_t/v-a_{t1})^2}{8\pi}D_X
-\frac{2a_G^{}{\hat s}}{v^2}\left(1+\frac{3m_H^2}{{\hat s}-m_H^2}\right)
-\frac{8a_G^2{\hat s}}{v^2},\\
{\mathcal M}^{+-}&={\mathcal M}^{-+}
= \frac{\alpha_s(\sqrt2m_t/v-a_{t1})^2}{8\pi}D_Y
-\frac{4a_G^2({\hat t}\,{\hat u}-m_H^2)}{v^2}\left(\frac1{\hat t}
+\frac1{\hat u}\right),
\end{align}
where ${\hat s}$, ${\hat t}$, and ${\hat u}$ are the Mandelstam variables
and the box functions $D_X$ and $D_Y$ are defined by
    \begin{align}
    D_X
    &= 4\left\{2+4\,m_t^2\,C_0({\hat s})
    -m_t^2\left({\hat s}+2\,m_H^2-8m_t^2\right)\left(D_0^{123}
    +D_0^{213}+D_0^{132}\right) \right.\nonumber \\
    &\quad +\frac{m_H^2-4\,m_t^2}{{\hat s}}
    \left[2\left({\hat t}-m_H^2\right)C_0^t
    +2\left({\hat u}-m_H^2\right)C_0^u
    \left.-\left({\hat t}\,{\hat u}-m_H^4\right)D_0^{132}
    \right]\right\},\\
    D_Y
    &= -2\left\{
    -{\hat s}\,{\hat t}\,D_0^{123}-{\hat s}\,{\hat u}\,D_0^{213}
	+2\,{\hat s}\,C_0({\hat s}
    +2\left({\hat t}-m_H^2\right)C_0^t
    +2\left({\hat u}-m_H^2\right)C_0^u \right. \nonumber \\
    &\quad +\left({\hat s}-2\,m_H^2+8m_t^2\right)
	\left[-2\,m_t^2\left(D_0^{123}
    +D_0^{213}+D_0^{132}\right)+2\,C_0^s\right.
    \nonumber \\
    &\quad\left.+\frac1{{\hat t}\,{\hat u}-m_H^4}
    \left(-{\hat s}\,{\hat t}^2\,D_0^{123}-{\hat s}\,{\hat u}^2\,D_0^{213}
    -{\hat s}\,\left({\hat s}-2\,m_H^2\right)C_0({\hat s})
    \right.\right.\nonumber \\
    &\qquad \left.\left.\left.-{\hat s}\left({\hat s}-4\,m_H^2\right)C_0^s
	+2\,{\hat t}\left({\hat t}-m_H^2\right)C_0^t
    +2\,{\hat u}\left({\hat u}-m_H^2\right)C_0^u\right)\right]\right\},
    \end{align}
where the scalar loop integrals are listed here:
    \begin{align}
    C_0^t=& C_0(0,{\hat t},m_H^2,m_t^2,m_t^2,m_t^2),\\
    C_0^u=& C_0(0,{\hat u},m_H^2,m_t^2,m_t^2,m_t^2),\\
    C_0^s=& C_0({\hat s},m_H^2,m_H^2,m_t^2,m_t^2,m_t^2),\\
    D_0^{123}=& D_0(0,0,m_H^2,m_H^2,{\hat s},{\hat t},m_t^2,m_t^2,m_t^2,m_t^2),\\
    D_0^{213}=& D_0(0,0,m_H^2,m_H^2,{\hat s},{\hat u},m_t^2,m_t^2,m_t^2,m_t^2),\\
    D_0^{132}=& D_0(0,m_H^2,0,m_H^2,{\hat u},{\hat t},m_t^2,m_t^2,m_t^2,m_t^2).
    \end{align}
The NLO correction for the Higgs boson pair production
in the gluon fusion mechanism has been studied
in the limit of the heavy top-quark mass~\cite{ggHHNLO}.
The effect increases the cross section to $1.9$--$2$
times that at the leading order. We here take
$K^\text{NLO}_{gg\to HH}= 1.9$ as the NLO correction to the
$gg\to HH$ process.

\section{Feynman Rules}
When we introduce the dimension-six genuine Higgs operators
${\mathcal O}_{\Phi1}^{}$ and ${\mathcal O}_{\Phi2}^{}$,
the Feynman rules for the Higgs self-interaction are given by~\cite{Barger:2003rs}
\begin{align}
HHH:&-6\,i\,v\,Z_{\Phi1}^3\left(Z_{\Phi1}^{-2}\frac{m_H^2}{2v^2}
+\frac{a_{\Phi1}^{}}{3v^2}\sum_{j<k}^3p_j\cdot p_k+\frac{a_{\Phi2}^{}}3\right)\\
HHHH:&-6\,i\,Z_{\Phi1}^4\left(Z_{\Phi1}^{-2}\frac{m_H^2}{2v^2}
+\frac{a_{\Phi1}^{}}{3v^2}\sum_{j<k}^4p_j\cdot p_k+2a_{\Phi2}^{}\right).
\end{align}
In the SM with the fermionic dimension-six operator ${\mathcal O}_{t1}^{}$,
the effective top--Higgs interactions are obtained as
\begin{align}
\overline{t}tH:& -i\,Z_{\Phi1}^{}\left(\frac{m_t}{v}-\frac{a_{t1}^{}}{\sqrt2}\right)\\
\overline{t}tHH:&  i\,Z_{\Phi1}^2\frac{3a_{t1}^{}}{\sqrt2v},
\end{align}
where the wave function renormalization due to the genuine Higgs operator
${\mathcal O}_{\Phi1}^{}$ is also taken into account.
The gluonic operator ${\mathcal O}_G^{}$ can induce gluon--Higgs vertices
at tree level as
\begin{align}
G_\mu^A(p_1)G_\nu^B(p_2)H:& -2i\,Z_{\Phi1}^{}\frac{a_G^{}}{v}p_1\cdot p_2
\left(g_{\mu\nu}-\frac{{p_2}_\mu{p_1}_\nu}{p_1\cdot p_2}\right)\delta^{AB}_{}\\
G_\mu^A(p_1)G_\nu^B(p_2)HH:& -2i\,Z_{\Phi1}^2\frac{a_G^{}}{v^2}p_1\cdot p_2
\left(g_{\mu\nu}-\frac{{p_2}_\mu{p_1}_\nu}{p_1\cdot p_2}\right)\delta^{AB}_{}.
\end{align}


\end{document}